\documentclass[twocolumn,superscriptaddress,epsf,psfig]{revtex4}
\usepackage{longtable}
\usepackage{amssymb}
\usepackage{epsfig}
\DeclareMathAlphabet{\pazocal}{OMS}{zplm}{m}{n}            
\usepackage{graphicx}
\usepackage{color}
\usepackage{bm}
\usepackage[normalem]{ulem}     
\usepackage{soul}
\usepackage{amsmath}
\usepackage{braket} 
\usepackage{rotating}
\usepackage{multirow} 
\DeclareMathAlphabet{\pazocal}{OMS}{zplm}{m}{n}      
\usepackage{graphicx}
\usepackage{hyperref}
\usepackage[toc,page]{appendix}

\begin{document}


\title{\textbf{Giant topological Hall effect in epitaxial Ni$_{80}$Fe$_{20}$/La$_{0.65}$Sr$_{0.35}$MnO$_3$ thin film heterostructures} 
}%

\author{Kusampal Yadav}
\affiliation{School of Physical Sciences, Indian Association for the Cultivation of Science,2A \(\&\)\ 2B Raja S. C. Mullick Road, Kolkata 700032, India}

\author{Dilruba Hasina}
\affiliation{School of Physical Sciences, Indian Association for the Cultivation of Science,2A \(\&\)\ 2B Raja S. C. Mullick Road, Kolkata 700032, India}
\author{Nasiruddin Mondal}
\affiliation{School of Physical Sciences, Indian Association for the Cultivation of Science,2A \(\&\)\ 2B Raja S. C. Mullick Road, Kolkata 700032, India}

\author{Sayantika Bhowal}
\email{sbhowal@iitb.ac.in}
\affiliation{Department of Physics, Indian Institute of Technology Bombay, Powai, Mumbai 400076, India}

\author{Devajyoti Mukherjee}
\email{sspdm@iacs.res.in}
\affiliation{School of Physical Sciences, Indian Association for the Cultivation of Science,2A \(\&\)\ 2B Raja S. C. Mullick Road, Kolkata 700032, India}

\date{\today}

\begin{abstract}

The emergence of new physical properties at the interfaces between complex oxides has always been of both fundamental and practical importance. Here, we report the observation of a giant topological Hall resistivity of $\sim 2.8 \mu \Omega$ \text{cm} at room temperature in an epitaxial thin-film heterostructure of permalloy (Py, Ni$_{80}$Fe$_{20}$) and the half-metallic ferromagnet La$_{0.65}$Sr$_{0.35}$MnO$_3$ (LSMO). This large magnitude of the topological Hall effect in the Py/LSMO heterostructure, compared to a single-layer Py thin film, is attributed to the optimized combination of ferromagnetism in LSMO and the strong spin-orbit-coupling-driven Rashba interaction at the interface. The introduction of a ferroelectric BaTiO$_3$ (BTO) sandwich layer in the Py/LSMO heterostructure also leads to an enhanced topological Hall resistivity compared to the single-layer Py thin film. Interestingly, magnetic force microscopy measurements reveal skyrmion-like features, suggesting the origin of the topological Hall effect. Our theoretical model calculations for the skyrmion lattice further indicate that the Rashba interaction, driven by the broken inversion symmetry in the Py/LSMO films, can account for the observed changes in the topological Hall effect at the interface. Our work opens the door for the potential use of Py/LSMO thin films in spintronic applications.  
\end{abstract}

                         
\maketitle

\section{Introduction}\label{intro}

A diverse range of electronic and magnetic properties have been observed at the interfaces and surfaces of complex oxides, unveiling novel phenomena absent in their bulk forms \cite{Hwang2012}. For instance, ferromagnetism can emerge at interfaces between materials that are not ferromagnetic in their bulk forms \cite{Takahashi2001, Nichols2016, BhowalSatpathy2019, Bhowal2019}, or even at interfaces between nonmagnetic bulk materials \cite{Ohtomo2004}. Similarly, bulk antiferromagnets can exhibit surface magnetism \cite{Belashchenko2010, Weber2023} and even multiferroicity \cite{Bhowal2024} driven by the bulk magnetic multipolar order combined with the inversion symmetry breaking at their surfaces. Exotic chiral magnetic textures including magnetic skyrmions have also been stabilized at oxide interfaces, enabled by strong spin-orbit coupling and broken inversion symmetry \cite{Matsuno2016, Ohuchi2018, Soumyanarayanan2016, Vistoli2019, Skoropata2020, natureLSMOSIOYoo}.

These emergent phenomena result from the interplay of competing interactions—exchange coupling, spin-orbit effects, and lattice distortions—alongside structural reconstructions, charge transfer, and symmetry breaking at the interfaces \cite{Ahn2006, Middey2014, Okamoto2017, Belabbes2016}. The flexibility of oxide materials, enhanced through epitaxial strain, chemical doping, or layer thickness control, enables precise engineering and manipulation of these properties. This makes oxide interfaces a versatile platform for exploring novel magnetic states and functionalities, with promising applications in next-generation spintronics and memory technologies \cite{Xiao2011, Matsuno2017}.

Here, we report an unusual behavior in the anomalous Hall effect (AHE) in the strained epitaxial thin films of permalloy (Py) interfaced with the half-metallic ferromagnet ${\rm La}_{0.65}{\rm Sr}_{0.35}{\rm MnO}_3$ (LSMO), suggesting the presence of a topological Hall effect (THE), in the Py/LSMO heterostructure which is about 5 times larger than that in a single-layer Py thin film. The introduction of a ferroelectric BaTiO$_3$ sandwich layer in the Py/LSMO hetero-structure also leads to enhanced THE as compared to the single-layer Py thin film due to the Rasba effect at the interface.

Permalloy (Py) is a magnetic alloy composed of approximately 80\% nickel and 20\% iron (${\rm Ni}_{80}{\rm Fe}_{20}$), known for its high Curie temperature and low spin polarization \cite{Huang2020}. It has been extensively studied, particularly for its large magnetic permeability, a property identified nearly a century ago \cite{Arnold1923}. Recent studies have pointed out the emergence of exotic spin textures at the interface between Py and magnetically doped topological insulators \cite{Zhang2018}.
In contrast, LSMO is a mixed-valence ferromagnetic perovskite distinguished by its high Curie temperature and near-complete spin polarization at the Fermi level \cite{Tokura1999}. LSMO has gained significant attention due to its complex interplay of charge, spin, and lattice interactions, which result in a rich and diverse phase diagram \cite{Coey1999}. Recent studies have also demonstrated the emergence of THE in heterostructures combining LSMO with materials like SrIrO$_3$ \cite{Li2019}.

Previous investigations of Py/LSMO/PMN-PT heterostructures have revealed phenomena like spin pumping and Rashba-type charge-to-spin conversion \cite{Zhao2023}, while the Py/LSMO interface itself has been analyzed for its magnetic, magnetoresistance, and magnetotransport properties \cite{Bergenti2018, Ruotolo2006}. These findings motivate further exploration of Hall transport properties in such heterostructures, forming the basis of this study.

Here, we report the growth of epitaxial thin films of Py, Py/LSMO, and Py/BTO/LSMO with a [100] orientation on single-crystal MgO (100) substrates using the pulsed laser deposition (PLD) method. Detailed X-ray diffraction analyses revealed residual strain in the epitaxial films due to the lattice mismatch with underlying substrates inducing tetragonal distortions in the cubic Py unit cell. Hall resistance measurements confirmed the presence of the THE across all heterostructures, with a significant enhancement in the topological Hall resistance observed in Py/LSMO and Py/BTO/LSMO films compared to single-layer Py films. 

This enhancement suggests the emergence of non-collinear spin textures, potentially including skyrmion-like topological spin configurations at the heterostructure interfaces. Field-dependent magnetic force microscopy reveals the presence of skyrmion-like non-coplanar spin textures of the heterostructures, explaining the origin of the observed THE. These phenomena are likely driven by the interfacial Rashba interaction caused by inversion symmetry breaking, combined with strong spin-orbit coupling. This is also supported by our theoretical modeling of a skyrmion lattice, which demonstrates that the magnitude of the topological Hall conductivity can be modulated by the strength of the Rashba interaction.  Our findings underscore the crucial role of interface effects in driving the THE at Py/LSMO interfaces and highlight the potential of such heterostructures for further exploration of spin textures and spintronic applications.

The manuscript is organized as follows: Section~\ref{methods} outlines the experimental techniques and provides a detailed description of the theoretical model employed in this study. Section~\ref{result} presents the results, including discussions on the crystallinity of the grown thin films, magnetization measurements, magneto-transport data, magnetic force microscopy measurements, and theoretical calculations. Finally, Section~\ref{summary} summarizes the findings and suggests potential directions for future exploration.

\section{Methodology}
In this section, we discuss the experimental techniques and the theoretical models employed in the present work.

\subsection{Experimental Methods} \label{methods}

High-purity (99.99 $\%$) commercial targets of permalloy (Py, Ni$_{80}$Fe$_{20}$), La$_{0.65}$Sr$_{0.35}$MnO$_3$ (LSMO) and BaTiO$_3$ (BTO) (Kurt J. Lesker Company) were used in this study. Py, Py/LSMO, and Py/BTO/LSMO thin film heterostructures were grown on single-crystal MgO (100) substrates using combined pulsed laser deposition (PLD; Neocera Pioneer 120 Advanced) and DC magnetron sputtering systems (Minilab S80A from Moorfield Nanotechnology). 
Initially, single-layer LSMO and BTO/LSMO bilayers were grown on 5 cm $\times$ 5 cm MgO (100) substrates using the PLD technique. Briefly, LSMO and BTO targets were sequentially ablated using a KrF excimer laser ($\lambda =$ 248 nm, frequency = 5 Hz, fluences = 3 J/cm$^2$) inside a deposition chamber equipped with a multi-target carousel allowing the in-situ deposition of multilayers with clean interfaces. A distance of 5 cm was maintained between the substrate and the targets during the depositions. Before growing the LSMO layers, the MgO substrates were annealed inside the PLD chamber at 800$^\circ$C under an ambient oxygen pressure $P(O_2)$ of 500 mTorr for 2 hrs. An initial layer of LSMO (thickness $\approx 70$ nm) was deposited at 800$^\circ$C under $P(O_2 )$ of 20 mTorr, followed by an ultrathin layer of BTO (thickness $\approx 10$ nm) at 750 $^\circ$C under a high $P(O_2)$ of 100 mTorr. After deposition, the samples were gradually cooled down to room temperature (approx. 4 hrs) under $P(O_2)$ of 100 mTorr. The as-deposited LSMO and BTO/LSMO thin films were immediately transferred to the sputtering chamber to deposit the Py top layers. The Py target was used to simultaneously deposit the single-layer Py thin film, Py/LSMO, and Py/BTO/LSMO heterostructures on MgO (100) substrates. The Py films were deposited at a base pressure of $10^{-7}$ mbar utilizing an input power of 55 W in Ar atmosphere (pressure $\sim 0.05$ mbar), whereas the substrates were rotated for homogeneous growth during deposition and heated to 450 $^\circ$C for high-quality crystallization. After deposition, the films were annealed for one hour at 450 $^\circ$C in a high vacuum (pressure of $10^{-7}$ mbar) within the deposition chamber before gradually cooling to ambient temperature. Throughout the deposition process, a quartz oscillator thickness monitor was used to track the thickness of the films in real time. The Py film thickness was fixed at 35 nm while the deposition rate was 1.4 \AA/s. All Py films were coated in situ with an ultrathin protective Au film ($\sim 2$ nm) to avoid surface oxidation. 

The crystallinity of the thin film heterostructures was measured using x-ray diffraction (XRD)-with a Rigaku Smart Lab 9 kW XG diffractometer provided with a 5-axis goniometer sample stage using collimated parallel beam Cu K$\alpha$ radiation ($\lambda= 1.5406$ \AA). The chemical composition of the films was determined using an X-ray photoelectron spectrometer (XPS, Omicron, model 1712-62-11) using a non-monochromatic Al K$\alpha$ (1486.7 eV) X-ray source that operates at 150 W (15 kV and 10 mA). 

The surface magnetic properties of the Py, Py/LSMO, and Py/BTO/LSMO heterostructures were investigated by measuring real-time magneto-optic Kerr effect (MOKE) ellipticity $(\epsilon_K)$ vs. magnetic field within a commercial MOKE setup operating in the longitudinal geometry. The details of the MOKE set-up are reported elsewhere ~\cite{MONDAL2022170118} (See Appendix \ref{In-situ Magneto-optic Kerr effect} for more details). The electrical transport and magnetic properties of the thin film samples were investigated using a Physical Property Measurement System (PPMS) (Quantum Design Inc., DynaCool 9T). Measurements included magnetization versus magnetic field M(H) hysteresis loops for both in-plane and out-of-plane configurations, as well as Hall resistivity, measured up to 5 T magnetic fields. For clarity, the presented resistivity data is limited to 2 T, while magnetization data is shown up to 3 T. The M(H) hysteresis loops were corrected by subtracting the diamagnetic contributions of the substrates.

The local magnetic properties of Py, Py/LSMO, and Py/BTO/LSMO heterostructures were examined using a Magnetic Force Microscope (MFM) equipped with a silicon probe (ASYMFM, Asylum Research) featuring pyramidal tips coated with a magnetic Co–Cr alloy film. Both morphological and magnetic force images were acquired simultaneously in this mode. To differentiate short-range topographic interactions from long-range magnetic signals, the measurements were performed in the ``tapping/lift" mode.  The probe was magnetized using a permanent magnet before performing the measurements. All MFM images, presented in this study, were captured with the tip magnetized perpendicular to the sample surface. During the first pass in tapping mode, topographic data was recorded, followed by lifting the tip by 50 nm to measure long-range magnetic signals. A variable field magnetic module integrated with the AFM setup was used to apply an in-plane external magnetic field (ranging from 0 to $\pm$0.2 T) to the sample surface, and MFM measurements were carried out under varying field strengths. All MFM data was analyzed using WSxM software \cite{Horcas}. 

\subsection{Theoretical Methods}

In order to understand if the Rashba interaction can explain the observed changes in the topological Hall conductivity (THC), we construct a generic tight-binding (TB) model for a skyrmion crystal on a square lattice. The TB model is given by
\cite{Hamamoto2015,BhowalSpaldin2022},
\begin{equation} \label{tb}
{\cal H}_s = t \sum_{\langle i, j \rangle} c_i^\dagger c_j -J_{\rm H} \sum_i \hat n_i \cdot (c_i^\dagger \vec \sigma c_i).
\end{equation}
Here $t$ and $J_{\rm H}$ are the nearest-neighbor hopping parameter and Hund's coupling respectively and the local magnetization vector $\hat n_i \equiv ( \sin\theta_i(r_i)\cos\phi_i(\alpha_i),  \sin\theta_i(r_i)  \sin\phi_i(\alpha_i),\cos\theta_i(r_i))$ describes the spin texture. Here the polar and azimuthal angles, viz., $\theta_i$ and $\phi_i$, depict the direction of the magnetization vector $\hat n_i$ at every lattice site $i$ of spatial coordinate $(r_i, \alpha_i)$. For a skyrmion texture, $\theta_i$ is only a function of $r$, and it varies from $\pi$ at the center of the spin texture to $0$ at the periphery of the spin texture. Accordingly, we can model it as $\theta_i=\pi (1-r_i/\lambda)$ \cite{Nagaosa-Tokura2013}. On the other hand, $\phi_i = m \alpha_i +\gamma$. Here the vorticity $m$ is an integer number and the helicity $\gamma$ can take different values. In the present work, we consider $m=1$ and $\gamma=0$, representing a N\'eel-type skyrmionic spin texture. We consider the adiabatic limit ($J_{\rm H} \gg t$) with $J_{\rm H}/t = 100$, at which the electronic spins are aligned to the direction of the local magnetization vector $\hat n_i$, describing the skyrmionic spin texture.

To investigate the effect of Rashba interaction on the THC, we consider the following Rashba term in the Hamiltonian, that occurs in the absence of the in-plane mirror symmetry, \cite{BhowalSatpathy2019, Kawano2023} 
\begin{equation} \label{HR}
{\cal H}_R = \alpha [\sigma_x \sin (k_ya)-\sigma_y \sin (k_xa)].
\end{equation}
Here $\sigma_x$ and $\sigma_y$ are the Pauli matrices, and $\alpha$ is the Rashba coefficient. In the small $k$ limit, the Hamiltonian in Eq. (\ref{HR}) reduces to ${\cal H}_R \propto (\sigma \times k)_z$, the well-known form of the Rashba interaction \cite{PekarRashba1964, Rashba1960, BychkovRashba}. We note that in real materials, the Rashba coefficient might differ for different bands \cite{Shanavas2014}. In the present case, however, for simplicity, we consider it to be identical for all the bands.

\section{Result and discussion}\label{result}

\subsection{Crystallinity and composition}
\begin{figure}
    \centering
    \includegraphics[width=1\linewidth]{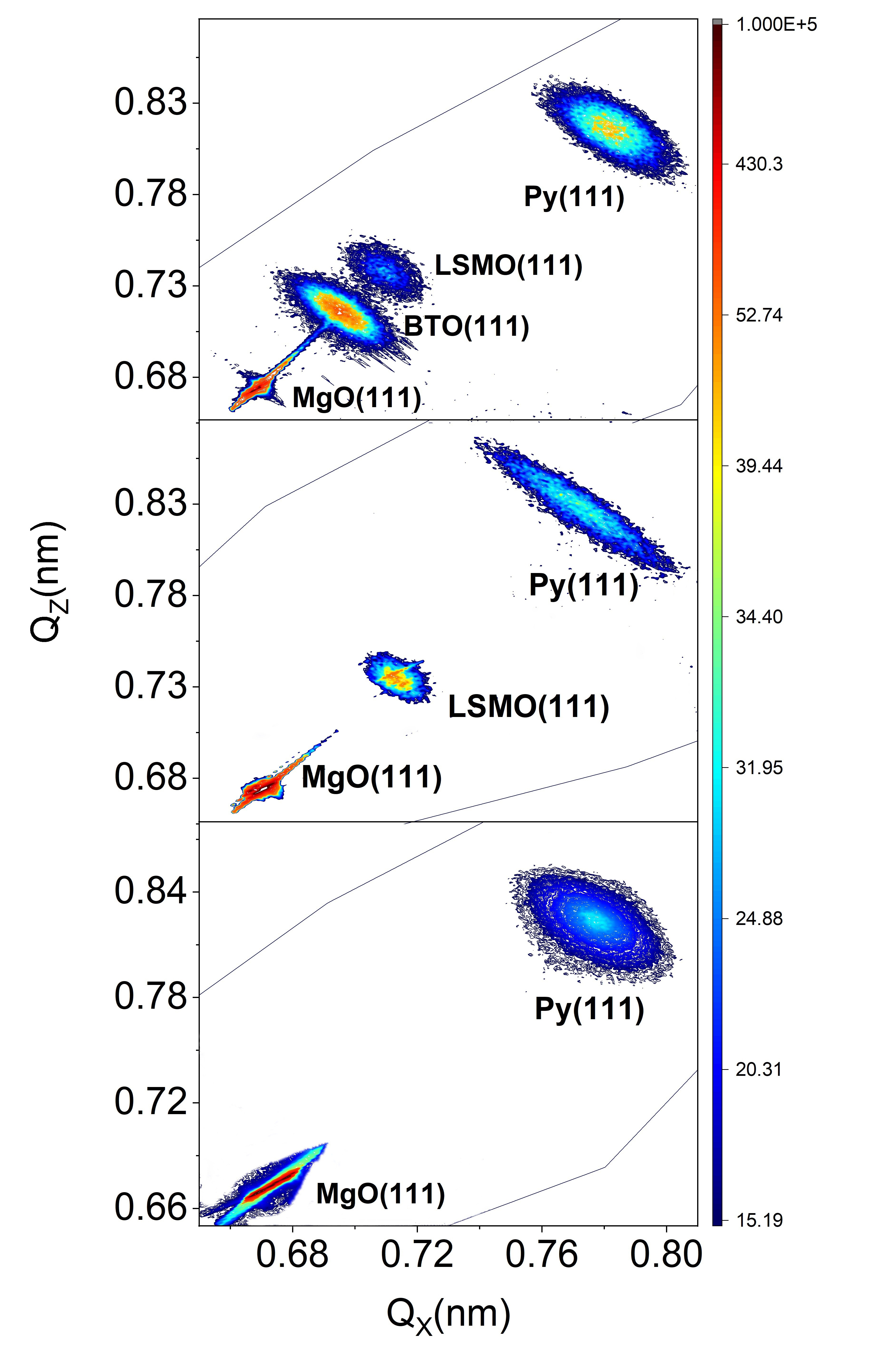}
    \caption{ Reciprocal space maps of Py ({\it bottom} panel), Py/LSMO ({\it middle} panel) and Py/BTO/LSMO ({\it top} panel) heterostructures performed around the (111) asymmetric plane of MgO single crystal substrate}
    \label{fig : 1}
\end{figure}

 The single-crystalline nature of the Py, Py/LSMO, and Py/BTO/LSMO heterostructures grown on single-crystalline MgO(100) substrates is confirmed through X-ray diffraction (XRD) high-resolution reciprocal space maps (RSMs) \cite{Jiang2019}, using asymmetrical reflections. Fig. \ref{fig : 1} shows the RSMs for the Py, Py/LSMO, and Py/BTO/LSMO heterostructures, measured around the MgO(111) plane.  
The RSMs reveal a single bright spot corresponding to the face-centered cubic (fcc) phase of Py(111) (lattice parameter $a = 3.55$ \AA, JCPDS No. 01-088-9591), the pseudo-cubic perovskite phase of LSMO(111) ($a = 3.88$ \AA, JCPDS No. 01-089-4461), and the tetragonal phase of BTO(111) ($a = 3.99$ \AA, $c = 4.03$ \AA, JCPDS No. 01-074-1957) near the MgO(111) substrate peak, confirming the epitaxial growth and crystallinity of the individual layers \cite{chatterjee2024interfacial} in all heterostructures.  

As expected, the Py(111) spot in the Py/MgO film is positioned far from the MgO(111) peak due to the large lattice mismatch of $\sim16 \%$. In contrast, for the Py/BTO/LSMO and Py/LSMO heterostructures, the Py(111) spot is much closer to the BTO and LSMO peaks due to their smaller lattice mismatches of $\sim 11 \%$ for Py/BTO/LSMO and $8.5\%$ for Py/LSMO, respectively (see Fig. \ref{fig : 1}).  The out-of-plane ($a_{\bot}$) and in-plane ($a_{\parallel}$) lattice parameters of the Py, LSMO, and BTO layers were calculated from the RSMs, along with the resulting tetragonal distortion $\left(\frac{a_{\bot}}{a_{\parallel}} - 1\right)$, out-of-plane strain $(\epsilon_{\bot})$, and in-plane strain $(\epsilon_{\parallel})$ relative to their bulk lattice parameters. These values are summarized in Table \ref{tab1}.

\begin{table*}[t]\label{tab1}
\caption{Out-of-plane ($a _{\bot}$) and in-plane ($a_\parallel$  ) lattice parameters, out-of-plane and in-plane epitaxial strains ($\epsilon_{\bot}$ and $\epsilon_{\parallel}$) and tetragonal distortion [$( \frac{a_{\bot} }{a_{\parallel}} )$ -1] for Py, LSMO and BTO layers of the Py, Py/LSMO, and Py/BTO/LSMO heterostructures.}
\begin{ruledtabular}
\begin{tabular}{lccccccc}
\textrm{\textbf{Sample}}&\textrm{$a _{\bot}$ (\AA)}&\textrm{$a_\parallel$(\AA)}&
\textrm{$\epsilon_{\bot}(\%)$}&\textrm{$\epsilon_{\parallel}(\%)$}&\textrm{Tetragonal distortion}&\\
&\multicolumn{2}{c}{}&\multicolumn{1}{c}{$( \frac{a_{\bot} }{a_{0}} )$ -1}&\multicolumn{1}{c}{$( \frac{a_{\parallel} }{a_{0}} )$ -1}&\multicolumn{1}{c}{$( \frac{a_{\bot} }{a_{\parallel}} )$ -1}\\
\hline
\colrule
\textbf{Py}&\\
Py layer & 3.45 ($\pm$0.02) & 3.62 ($\pm$0.02)  & -2.81 & 1.97& - 0.047\\
\textbf{Py/LSMO} &\\
Py-layer&3.47 (±0.03)&3.61 ($\pm$0.03)&-2.25&1.69&- 0.039\\
LSMO-layer&3.83 ($\pm$0.02) &3.95 ($\pm$0.01) &-1.29 &1.88 &- 0.03\\
\textbf{Py/BTO/LSMO}&\\
Py-layer&3.48 ($\pm$0.02)&3.59 ($\pm$0.03)&-1.97&1.13&-0.031\\
BTO-layer&4.06 ($\pm$0.01)&3.95 ($\pm$0.01)&1.65&-1.1&0.028\\
LSMO-layer &3.84 ($\pm$0.01) &3.97 ($\pm$0.01) & -1.03 &2.32 &- 0.033\\
\end{tabular}
\end{ruledtabular}
\end{table*}
XRD analysis reveals that the Py, LSMO, and BTO layers exhibit different strain states due to the epitaxial growth of the heterostructures on a lattice-mismatched MgO substrate. In the Py/MgO film, the Py layer experiences significant in-plane tensile strain $(\epsilon_{\parallel} = 1.97\%)$ and out-of-plane compressive strain $(\epsilon_{\bot} = -2.81\%)$, resulting in a pronounced tetragonal distortion of its unit cell. However, the strain values and corresponding distortion are reduced in the Py/LSMO and Py/BTO/LSMO heterostructures.  

Rocking curve measurements around the Py (200) crystallographic planes for these heterostructures yield peaks with narrow full-width-at-half-maximum (FWHM) values $(0.09^\circ \leq \Delta\omega \leq 0.15^\circ)$, indicating excellent in-plane orientation of the Py layers. Additionally, XRD azimuthal $(\phi)$ scans around the (111) asymmetric planes of the Py, BTO, and LSMO layers in the Py/BTO/LSMO thin film confirm their four-fold cubic symmetry and cube-on-cube epitaxial growth (see Appendix \ref{CRYSTALLINE INFORMATION} for more details).  

The chemical composition of the Py layer in the Py, Py/LSMO, and Py/BTO/LSMO heterostructures was verified using high-resolution XPS spectra. The core levels of Ni 2$p_{3/2}$ and Fe 2$p_{3/2}$ are shown in Appendix \ref{X-ray photoelectron spectroscopy(XPS)}. The calculated compositions closely match the nominal composition of Py (Ni$_{0.80\pm0.04}$Fe$_{0.20\pm0.03}$), indicating good stoichiometric growth of the heterostructures (see Appendix \ref{X-ray photoelectron spectroscopy(XPS)} for further details). 

\subsection{Magnetization and magneto-transport property measurement}
\begin{figure}[h]
    \centering
   \includegraphics[width=0.8 \linewidth]{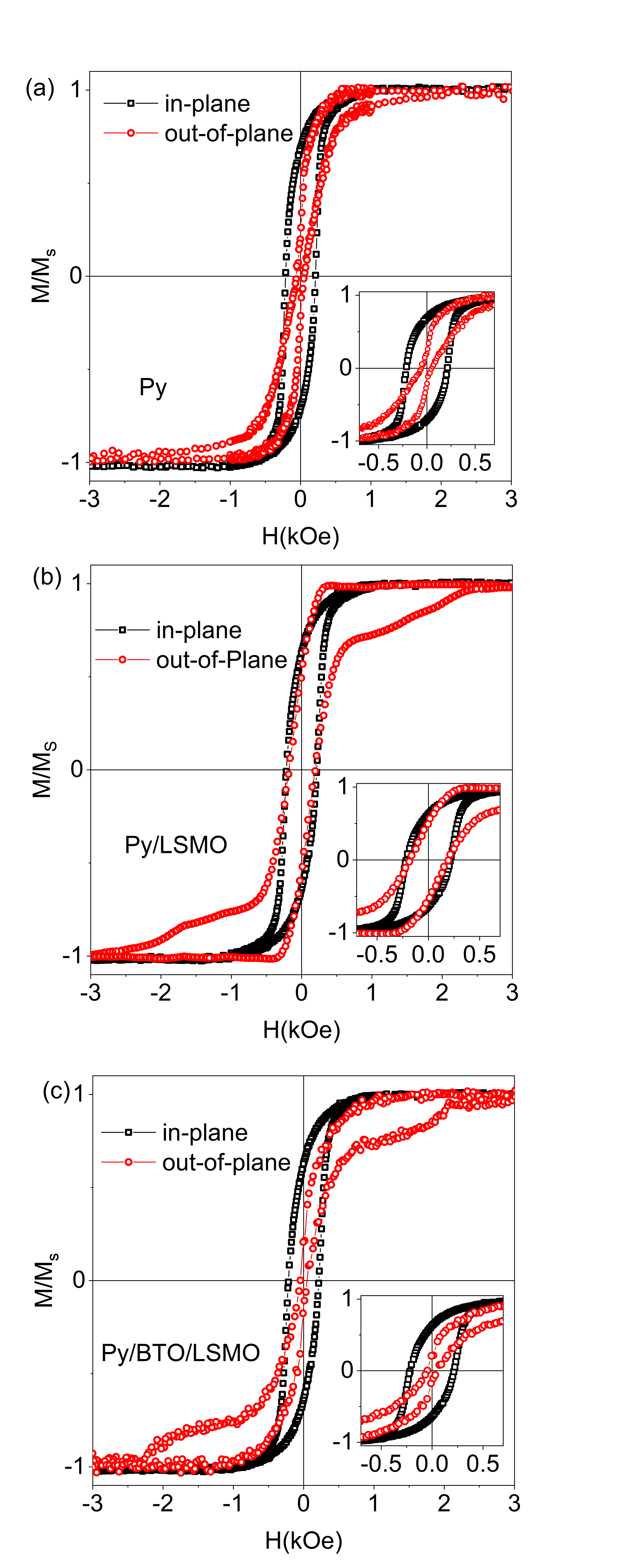}
    \caption{Normalized in- and out-of-plane magnetic M(H) hysteresis loops measured with the magnetic field applied parallel and perpendicular to the film plane for (a) Py, (b) Py/LSMO, and (c) Py/BTO/LSMO heterostructures, respectively. Insets show enlarged low-filed M(H) curves.}
    \label{fig:2}
\end{figure}

To explore the surface magnetic properties of these thin film heterostructures, longitudinal MOKE hysteresis loops were recorded. Our measurements show that all thin films display similar isotropic MOKE hysteresis loops which implies that the top Py layers in all the samples are the same (See Appendix \ref{In-situ Magneto-optic Kerr effect} for more details). To distinguish the surface magnetic properties of the thin films shown in Appendix \ref{In-situ Magneto-optic Kerr effect} from their volume magnetic properties, we have carried out measurements for the M(H) hysteresis loops for the thin films at room temperature, the results of which are shown in Fig. \ref{fig:2}. Distinct magnetic anisotropies are evidenced in the in-plane (IP) and out-of-plane (OOP) magnetic hysteresis loops shown in Figs. \ref{fig:2} (a)-(c) for the Py, Py/LSMO, and Py/BTO/LSMO hetero-structures, respectively. The magnetic parameters obtained from the hysteresis loops have been summarized in Table \ref{tab:table2}.
\begin{table*}[t]
\caption{\label{tab:table2}%
In-plane squareness in magnetic hysteresis loop $(M_{r\parallel}/M_{s\parallel})$, coercive field $(H_{C\parallel}(Oe))$ and out-of-plane squareness in magnetic hysteresis loop $(M_{r\bot} /M_{s\bot} )$, coercive field $(H_{C\bot} (Oe))$, uniaxial magnetic anisotropy constant $(K_{u})$ and maximum topological Hall resistivity$(\rho_{xy}^T(\mu\Omega-cm))$ for the Py, Py/LSMO and Py/BTO/LSMO heterostructures at room temperature.
}

\begin{ruledtabular}
\begin{tabular}{lcccccccc}

\textrm{\textbf{Sample}}&\textrm{$M_{r\parallel}/M_{s\parallel}$}&\textrm{$H_{C\parallel}(Oe)$}&\textrm{$M_{r\bot} /M_{s\bot} $}&\textrm{$H_{C\bot} (Oe)$}&\textrm{$K_{u}(erg/cm^3)$}&\textrm{$\rho_{xy}^T(\mu\Omega-cm)_{max}$}&\\
\hline
Py & 0.7 & 212.5 & 0.2 & 55 & $4.78\times10^5$ & 0.56\\
Py/LSMO & 0.63 & 217 & 0.5 & 194.2 & 1.97$\times10^5$ & 2.83\\
Py/BTO/LSMO & 0.64 & 215 & 0.18 & 23.5 & $2.60\times10^5$ & 2.60\\
\end{tabular}
\end{ruledtabular}
\end{table*}

From Fig. \ref{fig:2} (a) it is evident that the single-layer Py film exhibits uniaxial anisotropy with the easy axis of magnetization in the IP direction, evidenced by the high degree of squareness ($\frac {M_{r\parallel}}{M_{s\parallel}}$ $\sim$ 0.7) with large coercivity ($H_C=212.5$ Oe) in the IP hysteresis loop, while low squareness ($\frac {M_{r\bot}}{M_{s\bot}}$ $\sim$ 0.2) along with small coercivity ($H_C= 55$ Oe) in the OOP direction. From Fig. \ref{fig:2} (b), we find that the anisotropy in Py/LSMO is less than in Py thin film with comparable squareness and coercivity ($\frac {M_{r\parallel}}{M_{s\parallel}}$ $\sim$ 0.63, $H_C=217$ Oe) in the IP direction with that of the OOP direction ($\frac {M_{r\bot}}{M_{s\bot}}$ $\sim$ 0.5, $H_C=194.2$ Oe). From Fig. \ref{fig:2} (c) it is clear that the Py/BTO/LSMO film exhibits an uniaxial
anisotropy like the single-layer Py film, where the easy axis is along the IP direction and hard axis along the OOP direction. The uniaxial magnetic anisotropy constant (K$_u$) values obtained also show that Py has maximum anisotropy where Py/LSMO has minimum anisotropy (See Appendix \ref{Anisotropy calculation} for more details).The calculated values of $K_u$ are $4.78\times10^5 {\rm erg/cm}^2$ for Py, $1.97\times10^5 {\rm erg/cm}^2$ for Py/LSMO and $2.60\times10^5 {\rm erg/cm}^2$ for Py/BTO/LSMO. The lesser degrees of anisotropies in the Py/LSMO and Py/BTO/LSMO heterostructures as compared to the single Py thin film indicate the presence of exchange interactions across the interfaces of the heterostructures. 

Further from Fig. \ref{fig:2} (b), we can see that the OOP hysteresis loop exhibits features of a two-step magnetic hysteresis loop indicating two magnetic phase contributions; one behaving as a soft magnetic phase (i.e. Py) and another as a hard magnetic (LSMO). Such two-step magnetic hysteresis loops have also been observed in epitaxial SrRuO$_3$/La$_{0.42}$Ca$_{0.58}$MnO$_{3}$ \cite{D3NR02407E}, SrRuO$_{3}$/BiFeO$_{3}$ \cite{doi:10.1021/acsami.1c21703}, SrRuO$_{3}$/La$_{0.3}$Sr$_{0.7}$MnO$_{3}$ \cite{Rana2014} heterostructures, and associated to the presence of interfacial magnetic exchange interaction in these structures. Further, these step-like magnetic hysteresis loops have been considered as the origin of the observed THE in these heterostructures \cite{D3NR02407E, doi:10.1021/acsami.1c21703, LI2021127329, Rana2014, HE2022117619}. Again, from Fig. \ref{fig:2} (c), in the OOP hysteresis loop for Py/BTO/LSMO, we can observe a two-step hysteresis loop similar to the Py/LSMO heterostructure.  \cite{doi:10.1021/acsami.1c21703, Ziese}.

From Fig. \ref{fig:2} (b) and (c), it is observed that the OOP coercive field in Py/LSMO is larger than in Py/BTO/LSMO heterostructure. This implies the presence of exchange coupling between Py and LSMO layers in Py/LSMO heterostructure \cite{D3NR02407E}. The similar values of OOP coercive fields for single-layer Py and Py/BTO/LSMO heterostructure confirm the absence of exchange coupling in Py/BTO/LSMO heterostructure \cite{D3NR02407E}. The presence of the sandwich BTO layer effectively decouples the magnetic exchange interaction between Py and LSMO layers in the Py/BTO/LSMO heterostructure. Magnetic exchange interaction in Py/LSMO can arise due to several modifications at the Py-LSMO interface, such as orbital hybridization and charge transfer across the interface \cite{LI2021127329,natureLSMOSIOYoo,doi:10.1021/acsomega.1c06529LaSrMnO3}. Specifically, electron transfer from Fe and Ni in Py to Mn in LSMO, via bridging oxygen anions, strengthens this exchange coupling, further stabilizing the interfacial magnetic interactions \cite{LI2021127329}. Additionally, this charge transfer generates an interfacial electric field \cite{PhysRevB.91.054419D.MukherjeePZTCFOLSMO}, which may enhance the THE in the Py/LSMO bilayer system, as we discuss later in section \ref{Model}. 
On the other hand, the presence of spontaneous polarization in BTO in Py/BTO/LSMO could potentially give rise to enhanced THE due to ferroelectric proximity effect \cite{Py/BTO}, similarly to the SrRuO$_\mathrm{3}$/BiFeO$_\mathrm{3}$ heterostructure  \cite{doi:10.1021/acsami.1c21703}.

\begin{figure*}[t]
    \centering
     \includegraphics[width=1 \linewidth]{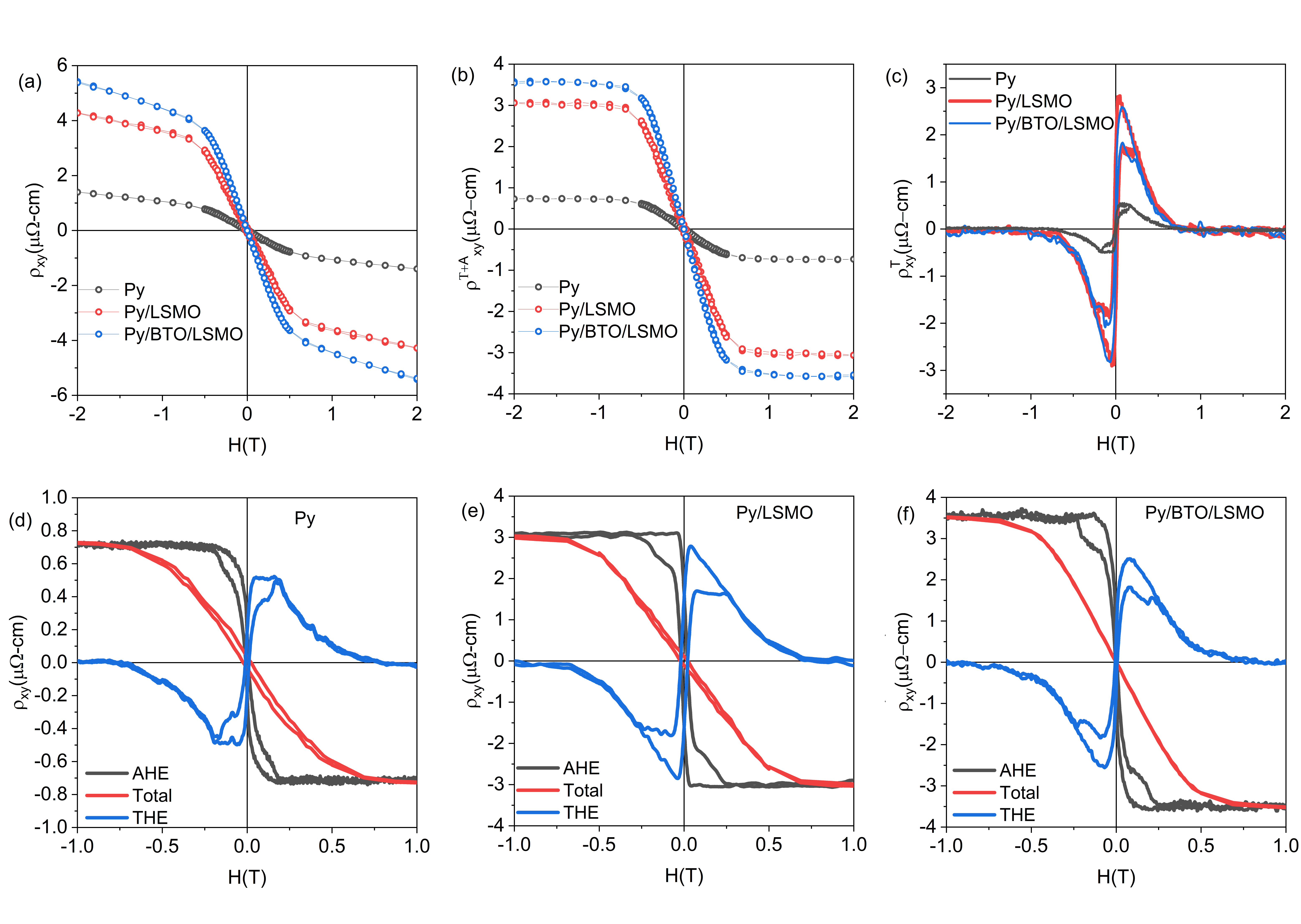}
    \caption{(a) Total Hall resistivity (b) combined anomalous and topological Hall resistivity, and (c) topological Hall resistivity curves in the magnetic field range of $\pm$ 2 T for the Py, Py/LSMO, and Py/BTO/LSMO heterostructures. (d, e, f) Field dependence of total(AHE+THE), AHE and THE resistivity in the range $\pm$ 1T for Py, Py/LSMO and Py/BTO/LSMO heterostructures, respectively.}
    \label{fig:THEdata}

\end{figure*}

Fig. \ref{fig:THEdata} shows the room temperature total Hall resistivity $(\rho_{xy})$ vs magnetic field curves for Py, Py/LSMO, and Py/BTO/LSMO heterostructures. The detailed calculation of the OHE, AHE, and THE resistivities from the total Hall resistivity have been shown in Appendix: \ref{Topological Hall calculation}. The carrier concentrations calculated from the ordinary Hall effect (OHE) using Fig \ref{fig:THEdata} (a) for the Py film is $1.88(\pm1)\times10^{23} {\rm cm}^{-3}$, for the Py/LSMO film is $1.02(\pm3)\times10^{23} {\rm cm}^{-3}$ and for the Py/BTO/LSMO film is $6.77(\pm5)\times10^{22}  {\rm cm}^{-3}$, respectively. The decrease in carrier concentration in the Py layers in the heterostructures as compared to the single-layer Py thin film clearly indicates the presence of charge transfer across the interfaces which could give rise to interfacial electric fields in the heterostructures. Fig. \ref{fig:THEdata} (b) shows the sum of the anomalous Hall resistivity and topological Hall resistivity, i.e. $\rho_{xy}^{A+T}=\rho_{xy} (H)-\rho_{xy}^0$ for Py, Py/LSMO and Py/BTO/LSMO heterostructures. From Fig. \ref{fig:THEdata} (b), we notice that Py/LSMO and Py/BTO/LSMO heterostructures show much higher values of  $\rho_{xy}^{A+T}$ as compared to single-layer Py film at all fields. This could be attributed to the presence of two magnetic phases and exchange coupling in the heterostructures giving rise to higher AHE \cite{D3NR02407E, doi:10.1021/acsami.1c21703, LI2021127329, Rana2014, HE2022117619}. From Fig.~\ref{fig:THEdata}(c), it can be seen that the maximum value of  $\rho_{xy}^{T}$ is higher for the Py/LSMO and Py/BTO/LSMO heterostructures as compared to that in the single-layer Py thin film.
From Figs.~\ref{fig:THEdata} (b) and (c), it is observed that the topological Hall resistivity can withstand up to a magnetic field of 0.8 T for all samples, where anomalous Hall resistivity is saturated at a field of nearly 0.2 T. The maximum value of topological Hall resistivity ($\rho_{xy}^T )_{max} $ for Py/LSMO and Py/BTO/LSMO thin films are 2.83 $\mu \Omega$ cm and 2.60 $\mu \Omega$ cm, respectively, that are significantly higher than the single-layer Py film (0.56 $\mu \Omega$ cm ). Such high values of THE resistivity have hitherto remained unobserved in Py/LSMO thin films. A comparative summary of the reported materials is listed in table \ref{tab:table3}.

\begin{table*}[t]
\caption{\label{tab:table3}%
Summary of values reported in literature for the maximum topological Hall resistivity $(\rho_{xy}^T(\mu\Omega{\rm cm}))_{max}$ measured at specific temperature (T(K)) and the value of the magnetic field where the maximum topological Hall resistivity is observed $(H_{\rho max}(T))$ for different bulk materials, single crystals and epitaxial thin films with the observed type of spin structure.
}
\begin{ruledtabular}
\begin{tabular}{cccccc}
\textrm{\textbf{Material}}&\textrm{$\rho_{xy}^T(\mu\Omega{\rm cm})_{max}$}&\textrm{$T(K)$}&\textrm{$H_{\rho_{max}}(T)$}&\textrm{Spin structure}&\textrm{Reference}\\
\hline
\colrule
\textbf{Bulk}\\
\\
MnGe & $\sim$0.3 & 160 K &$\sim$0.06 T & Spin chirality (Skyrmions) &  \cite{PhysRevLett.106.156603Kanazawa} \\
Mn$_{2}$PtSn (Heusler alloy) & 1.53 & 150 K & 0.6 T & Non-collinear spin & \cite{LIU2018122Z.Hliu} \\ 
EuCuAs(single crystal) &7.4 &13 K &&Non trivial spin &\cite{doi:10.1021/jacs.3c04249EuCuAs}\\
MnBi$_{\mathrm{4}}$Te$_{\mathrm{7}}$ (single crystal)&7&2 K&0.5 T&Noncoplanar spin&\cite{doi:10.1021/acs.chemmater.1c02625MnBi4Te7}\\
GdCoC$_{\mathrm{2}}$ &0.23&3 K&0.6 T&Spin chirality (Skyrmions)&\cite{10.1063/5.0160745GdCoC2}\\
 MnNiGa&0.15&200 K&&Biskyrmions&\cite{https://doi.org/10.1002/adma.201600889MnNiGa}\\
{Mn}$_{2-x}${Zn}$_x$Sb(single crystal) & $\sim$2&250 K&$\sim$0.07 T&Geometric
frustration(Skyrmions) & \cite{PhysRevB.104.174419Nabi} \\ 
\textbf{Films}\\
\\
FeGe & $\sim$0.06 & 50 K& $\sim0.2$ T& Chiral cone spin and Symerion(interface DMI) &\cite{PhysRevMaterials.2.041401Chiral} \\ 
FeGe&1& 300 K &$\sim$0.05 T & Skyrmions &\cite{PhysRevLett.118.027201Gallagher} \\ 
FeGe&2.34&330 K&0.4 T&Non-collinear spin(Skyrmions)&\cite{PhysRevB.101.220405Budhathoki}\\
 Pt/Co/Ta   & $\sim$0.15&300 K& $\sim0.9$ T & Nontrivial skyrmions&\cite{PhysRevB.97.174419Hemin}  \\ 
SrRuO$_{\mathrm{0.3}}$/SrIrO$_{\mathrm{0.3}}$ & 0.2&80 K&$\sim$0.06 T&Skyrmions(interface DMI) & \cite{Matsuno2016} \\
LaMnO$_{\mathrm{3}}$/SrIrO$_{\mathrm{0.3}}$&75&10 K&&Interface Skyrmions&\cite{Skoropata2020}\\
La$_{0.7}$Sr$_{0.3}$MnO$_{3}$/SrIrO$_{3}$ & $\sim1$ & 200 K&&Non-collinear spin(Skyrmions)& \cite{Li2019} \\ 
Ca$_{\mathrm{0.99}}$Ce$_{\mathrm{0.01}}$MnO$_{\mathrm{3}}$&120&20 K&$\sim$4 T& Non trivial spins(magnetic bubbles)&\cite{Vistoli2019}\\
EuO&12&50 K&$\sim1.5$ T& noncoplanar
spin (2D skyrmions)&\cite{PhysRevB.91.245115EuO}\\

Cr$_{\mathrm{5}}$Te$_{\mathrm{6}}$&1.6&90 K&$\sim0.9$ T& Non coplanar spin&\cite{https://doi.org/10.1002/adfm.202302984Cr5Te6}\\ 

MgO/CoFeB/Ta&0.77&50 K&$\sim0.75$ T&Skyrmions(interfacial DMI)&\cite{CoFeB}\\
 Pd/CoZr/MgO&0.52&100 K&&Interfacial DMI&\cite{10.1063/5.0127474Pd/CoZr/MgO}\\

 NdCuCo&0.16&125 K&0.5 T&Biskymerion \& Non trivial spins texture &\cite{10.1063/5.0128572NdCuCo}\\
 MnCoAl&0.058&300 K& &Skyrmions(Magnetic bubbles)&\cite{PhysRevB.104.064409MnCoAl}\\
Fe$_{\mathrm{0.7}}$Co$_{\mathrm{0.3}}$Si&0.82&5 K&0.65 mT&Skyrmions&\cite{porter2013gianttopologicalhalleffectFeCoSi}\\
Fe-Co-Ni-Mn&1.9&300 K&$\sim0.1$ T&Non-coplanar spin&\cite{JihaoYuAM}\\
Py&0.56&300 K&0.1 T& Non coplanar spin &This work\\
Py/LSMO  & 2.83&300 K&0.05 T&Non coplanar spin & This work\\
Py/BTO/LSMO&2.60&300 K&0.08 T&Non coplanar spin& This work

\end{tabular}
\end{ruledtabular}
\end{table*}

 The nominal THE observed in the single-layered Py film $((\rho_{xy}^T)_{\rm max}$ = 0.56 $\mu\Omega·$ cm) possibly arises from the tetragonal distortion of the cubic Py unit cell due to in-plane tensile strain \ref{tab1}  \cite{PhysRevB.109.224407Pystrain} that breaks inversion symmetry and possibly generates Dzyaloshinskii-Moriya interactions (DMI)\cite{PhysRevB.110.014402DMINICu, PhysRevLett.101.197204PyDMI,SROepitaxy} within the spins. However, the significant enhancement of the THE in the Py/LSMO heterostructure $((\rho_{xy}^T )_{max} = $2.83 $\mu\Omega·$ cm) is most likely due to the effect of the built-in electric field at the Py-LSMO interface and the presence of magnetic exchange coupling between Py and LSMO layers. The built-in electric field gives rise to Rashba-type spin-orbit coupling and interfacial DMI in Py/LSMO heterostructure \cite{WOS:000314741200031PyLSMO}. The changes in the oxidation states of Ni and Fe observed in the XPS spectra of the Py/LSMO heterostructure further corroborate the flow of charge carriers between the Py and LSMO layers. This charge transfer is likely driven by the formation of a potential gradient across the heterostructure, resulting in a built-in electric field. In the case of the Py/BTO/LSMO, the spontaneous polarization of the BTO sandwich layer generates an intrinsic built-in electric field at the Py-BTO interface, which possibly breaks the inversion symmetry of the Py structure, giving rise to an interfacial DMI \cite{Py/BTO,doi:10.1021/acsami.1c21703}.  

Fig. \ref{fig:THEdata} (d)-(f) show that the magnetic field dependence of the total Hall resistivity, the $\rho_{xy}^A$  and the $\rho_{xy}^T$  at low magnetic field range of $\pm$1 T for the Py, Py/LSMO and Py/BTO/LSMO heterostructures. From the figure, we can see that both the $\rho_{xy}^A$ and the $\rho_{xy}^T$ curves exhibit distinct hysteresis loops; however, the directions of their hysteresis loops are opposite. From the Figs.~\ref{fig:THEdata} (e) and (f), it is observed that the saturation values for the AHE resistivity are slightly higher in Py/BTO/LSMO (3.5 $\mu\Omega$ cm) as compared to that in Py/LSMO (3 $\mu\Omega$ cm) which could be associated to the higher value of the anisotropy constant in the Py/BTO/LSMO as compared to Py/LSMO (\ref{tab:table2}). On the other hand, the saturation values of THE resistivity is slightly higher in Py/LSMO as compared to that in Py/BTO/LSMO possibly due to the enhanced effect of the interfacial DMI in Py/LSMO heterostructure. It can be observed that all heterostructure shows remanent $\rho_{xy}^T$ at H=0 field i.e., 0.19 $\mu\Omega$ cm , 0.63$\mu\Omega$
cm and 1.55 $\mu\Omega$ cm for Py, Py/BTO/LSMO, and Py/LSMO respectively. We further note that the Py/LSMO heterostructure shows the highest remanent $\rho_{xy}^T$, 1.55 $\mu\Omega$ cm, which indicates stable skyrmion formation at zero field. This could be particularly advantageous for technological applications due to the stability of skyrmions under these conditions. \cite{PhysRevLett.118.027201Gallagher}. The temperature-dependent Hall resistivity measured from 250 K to 375 K also shows that the THE is stable for a wide range of temperatures in all heterostructures (See Appendix: \ref{Temperature dependent topological Hall resistivity} for more details).

\begin{figure}[h]
    \centering
\includegraphics[scale=0.22]{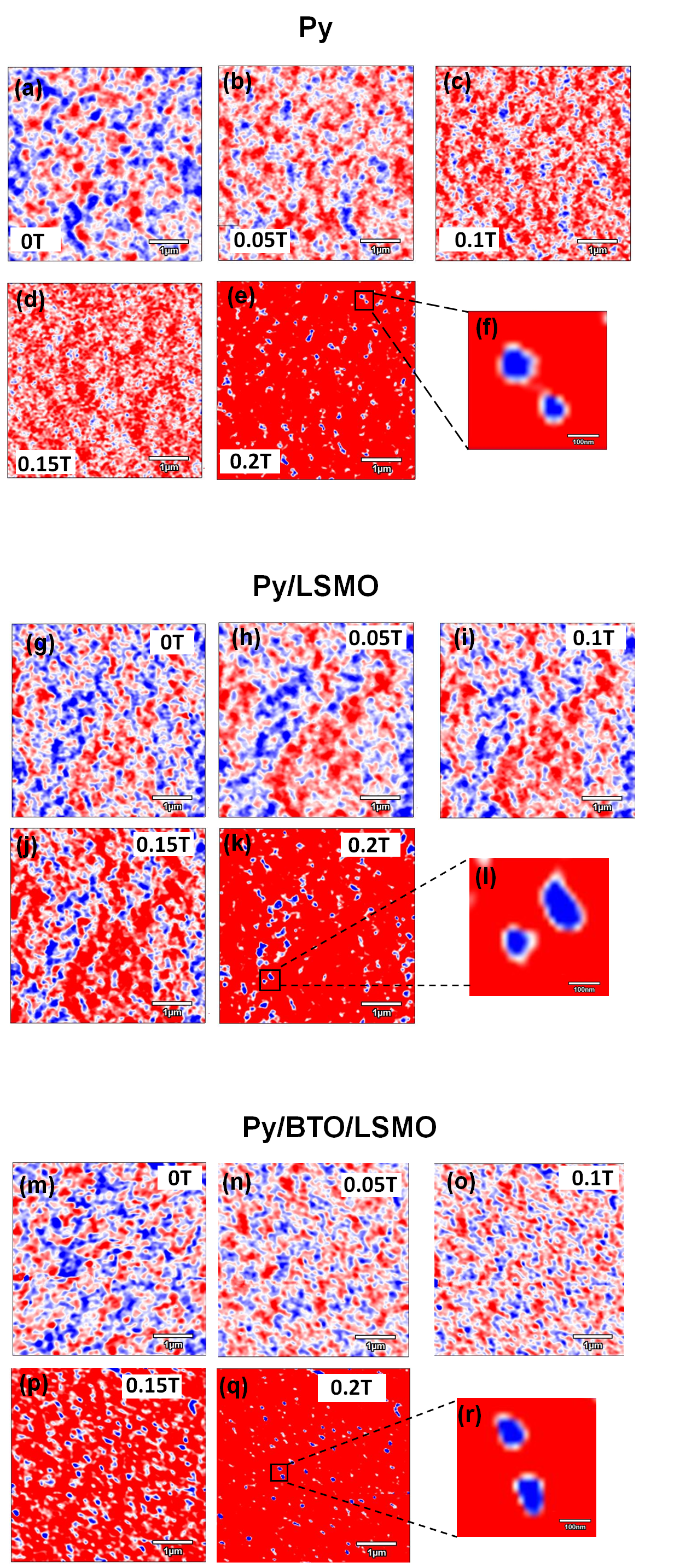}
    \caption{ MFM images of the top Py layers in (a-f) Py, (g-l) Py/LSMO, and (m-r) Py/BTO/LSMO captured at various fields from 0 to 0.2 T: (a), (g), (m) 0 T; (b), (h), (n) 0.05 T; (c), (i), (o) 0.1 T; (d), (j), (p) 0.15 T; (e), (k), (q) 0.2 T. The scale bar at the bottom of the images represents 1 $\mu m$. (f), (l), and (r) show a zoom-in view of the regions marked with black boxes in (e), (k), and (q), respectively. The scale bar for (f, l, r) represents 100 nm. Blue/red contrasts denote negative/positive magnetization regions along the z-axis.}
    \label{fig:MFM}
\end{figure}

To explore the presence of non-coplanar spin textures in the Py layers, field-dependent MFM imaging was performed at room temperature for all the thin films. Magnetic fields were systematically varied from 0 T to $\pm$0.2 T in increments of 0.05 T, applied perpendicular to the thin film surface in both positive and negative directions. Figs. \ref{fig:MFM} (a-f), (g-l), (m-r) shows the MFM images of the Py, Py/LSMO, and Py/BTO/LSMO heterostructures, respectively, captured at various fields from 0 to 0.2 T applied in the positive $z$ direction. The red and blue contrasts in the images correspond to the magnetization along the $+z$ and $-z$ directions, respectively. The white regions in the figures indicate regions of weak magnetic signals when the spins are oriented along the in-plane direction of the thin films. The corresponding MFM images captured at various fields applied in the negative direction are shown in Fig. \ref{fig:MFMsupplemetry} in Appendix : \ref{Magnetic Force Microscope_suppl (MFM)}. At 0 T the MFM images show similar magnetic domains in all the samples with distinct labyrinthine patterns characterized by worm-like structures and elongated stripes, as shown in Figs.~\ref{fig:MFM} (a), (g) and (m) for Py, Py/LSMO, and Py/BTO/LSMO heterostructures, respectively. As the field is increased gradually from 0.05 to 0.1 T, we can observe that the domains that are oriented along the direction of the field (i.e., red-colored domains) gradually grow in size as more and more spins are directed along the field direction until almost all the domains are red with few blue domains for the Py (Figs.~ \ref{fig:MFM} b-d), Py/LSMO (Figs.~ \ref{fig:MFM} h-j) and Py/BTO/LSMO (Figs.~ \ref{fig:MFM} n-p) heterostructures. However, at higher fields of 0.2 T (Figs.~ \ref{fig:MFM} e, k, q) when all the domains should have turned red, the images reveal the emergence of small, skyrmion-like structures, which are highlighted in the magnified views in Figs. \ref{fig:MFM} (f), (l), (r) of Py, Py/LSMO, and Py/BTO/LSMO heterostructures, respectively. The images clearly show the skyrmion-like features are irregular in shape and have distinct white boundaries separating the oppositely oriented spins. The average skyrmion sizes for Py, Py/LSMO, and Py/BTO/LSMO are approximately 55 nm, 137 nm, and 75 nm, respectively, as evident in the magnified images of Figs. \ref{fig:MFM} (f), (l), and (r). Notably, the average size of the skyrmions in Py/LSMO is the largest, while that in single-layer Py is the smallest. The experimental observation of the skyrmion-like features in the field-dependent MFM analyses is consistent with the data on topological Hall resistivity in Fig. \ref{fig:THEdata} (d, e, and f).

\subsection{Results of the model calculations}\label{Model}

\begin{figure}[t]
\centering
\includegraphics[width=\columnwidth]{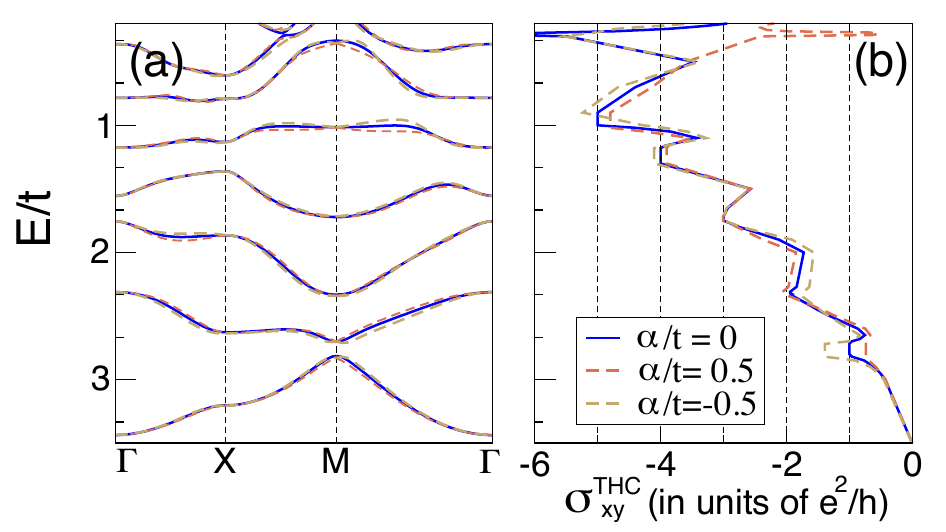}
 \caption{(a) Band structure of the skyrmion lattice along different high symmetry $k$ points for three different values of $\alpha/t$, viz., 0 and $\pm 0.5$. Here $\alpha$ is the Rashba coefficient. (b) The corresponding variation of the topological Hall conductivity $\sigma^{\rm THC}_{xy}$ as a function of $E/t$, where $E$ is the band energy and $t$ is the hopping parameter. 
 }
 \label{fig-model}
 \end{figure}

As stated before, we anticipate that the observed changes in the THE in the heterostructures originate from the interfacial Rashba interaction.
In order to, therefore, explicitly investigate the effect of Rashba interaction on the THC,  
we  compute the THC $\sigma^{\rm THC}_{xy}$ for the TB model ${\cal H} = {\cal H}_s + {\cal H}_R$ using the Kubo formula,
\begin{equation} \label{cond}
\sigma^{\rm THC}_{xy} = \frac{e^2}{h} \frac{2\pi}{S_A} \sum^{\text occ}_{n,\vec k} \Omega^z_{n} (\vec k). 
\end{equation}
Here the sum runs over the occupied part of the Brillouin zone (BZ) and $e, S_{\rm A},$ and $ n$ are the electronic charge, surface area of the skyrmion lattice, and the band index respectively. The $z$ component of the Berry curvature $\Omega^z_{n}$ for the $n$th band is given by,  
\begin{equation} \label{BC}
\Omega^z_{n} (\vec k) = 2~\text{Im} \sum_{m \ne n} \frac{\bra {n\vec k} \frac{\partial {\cal H}}{\partial k_x} \ket{m \vec k}\bra {m\vec k} \frac{\partial {\cal H}}{\partial k_y} \ket{n \vec k}}{(E_{n\vec k}-E_{m\vec k})^2}.
\end{equation}
Here $\ket{n \vec k}$ and $E_{n\vec k}$ are respectively the eigenstates and the energy eigenvalues of the TB model ${\cal H} $. The computed band structure of the Hamiltonian ${\cal H}$ for three different values of $\alpha/t$ is shown in Fig. \ref{fig-model} (a). The corresponding computed 
$\sigma^{\rm THC}_{xy}$ as a function of the Fermi energy is also shown in Fig. \ref{fig-model} (b). 

First, we analyze the case of $\alpha/t = 0$, at which ${\cal H} = {\cal H}_s$. In this case,
as seen from Fig. \ref{fig-model}, the computed THC is non-zero, and becomes quantized, i.e., an integer multiple of $e^2/h$ if the Fermi energy lies within the band gap. We note that the quantized THC is a characteristic of the skyrmion crystal. In reality, however, there can be the formation of isolated skyrmions at the heterostructure (as experimentally observed in the MFM images in Fig. \ref{fig:MFM}) rather than a lattice of 
skyrmions as considered in the model, which will lead to deviation from quantization. In this context, it is important to point out that the emergence of topological spin texture is not the only possibility to explain the observed anomaly in the anomalous Hall conductivity (AHC). Non-topological chiral spin textures as well as inhomogeneous electronic structures in
thin films may also lead to similar anomalies in the AHE \cite{Kimbell2022}. Hence,  confirmation of the existence of topological spin texture requires further investigations. Real-space imaging using Lorentz transmission electron microscopy \cite{Zhao2016, Tang2019, Birch2020}, for example, can be a decisive measurement for the confirmation of topological spin texture.

Next we analyze the case with $\alpha/t \ne 0$, i.e., including the effect of Rashba interaction. The mirror symmetry breaking at the interface leads to the Rashba interaction. The Rashba interaction is known to contribute to the stabilization of magnetic skyrmions \cite{Sahu2022}. In addition to that, if the skyrmionic spin texture is present, the Rashba interaction also affects the magnitude of the THC, as revealed by our calculations. As seen from Fig. \ref{fig-model} (b), the magnitude of the THC changes in the presence of the Rashba interaction. We correlate these changes in the THC to the corresponding changes in the band structure in the presence of the Rashba interaction. We notice that the increase or decrease in the value of the THC is, however, dependent on the detail of the electronic structure and the sign of the Rashba coefficient $\alpha$. Notably, as we see from Fig. \ref{fig-model} (b), the changes in the THC, induced by the Rashba interaction, are opposite for the opposite signs of the Rashba coefficient, suggesting a possible control of the THC using an external electric field. Overall, our model study shows that the presence of Rashba interaction can lead to changes in the magnitude of the THC and, therefore, could be a possible origin of the observed increase in the THE at the Py/LSMO and Py/BTO/LSMO interfaces as compared to single Py thin film. We hope our work will motivate future experiments using an external gate voltage to control the THC at the Py/LSMO interfaces of these heterostructures.   

\section{Summary and outlook}\label{summary}

In summary, we report a pronounced THE alongside an AHE in epitaxial Py/LSMO and Py/BTO/LSMO thin film heterostructures grown on MgO (100) substrates. The single-crystalline natures of the Py, Py/LSMO, and Py/BTO/LSMO thin films are confirmed using X-ray diffraction reciprocal space maps, which also reveal the presence of residual strain in the films. Field-dependent MFM analyses reveal the presence of skyrmion-like spin textures in all the Py thin films. Our numerical calculations based on a model Hamiltonian provide crucial insight into the observed THE at the interface. 

Our analysis addresses two significant observations: first, the anomaly in the AHC of Py thin films, suggesting the presence of topological Hall conductivity (THC), potentially arises from the emergent topological spin textures. Second, the further enhancement of THC in the Py/LSMO heterostructures, which our calculations attribute to interfacial Rashba interactions. These findings highlight the complex interplay of spin texture and interfacial effects in governing the Hall responses in such systems.

Further investigations are necessary to confirm the existence of topological spin textures in Py/LSMO heterostructures. Micro-magnetic simulations tailored to the Py/LSMO interface can also offer crucial insights into the nature of spin textures, whether topological or non-topological. Furthermore, the role of interfacial Rashba interactions could be explored through Hall measurements under external gate voltage, as suggested by our theoretical predictions. 

We hope this study inspires further experimental and theoretical work to explore spin textures and interfacial phenomena in similar systems, with potential implications for spintronic applications and beyond.

\begin{acknowledgments}
D.M. expresses gratitude for the financial support received from the Technical Research Center, Department of Science and Technology (DST), Government of India (Grant No. AI/1/62/IACS/2015) and the India Russia Joint Research Call, Department of Science and Technology (DST), Government of India (Grant No. DST/IC/RSF/2024/542). S.B. thanks National Supercomputing Mission for providing computing resources
of ‘PARAM Porul’ at NIT Trichy, implemented by C-DAC and supported by the Ministry
of Electronics and Information Technology (MeitY) and Department of Science and
Technology, Government of India, and acknowledges funding support from the Industrial Research and Consultancy Centre (IRCC) Seed Grant (RD/0523-IRCCSH0-018) and the INSPIRE research grant (project code RD/0124-DST0030-002). The authors acknowledge Prof. Tapobrata Som, Institute of Physics, Bhubaneswar for providing access to the MFM measurement facilities. 
\end{acknowledgments}

\appendix
\newpage
\section{CRYSTALLINE INFORMATION} \label{CRYSTALLINE INFORMATION}
\begin{figure}[h]
    \centering
    \includegraphics[width=1\linewidth]{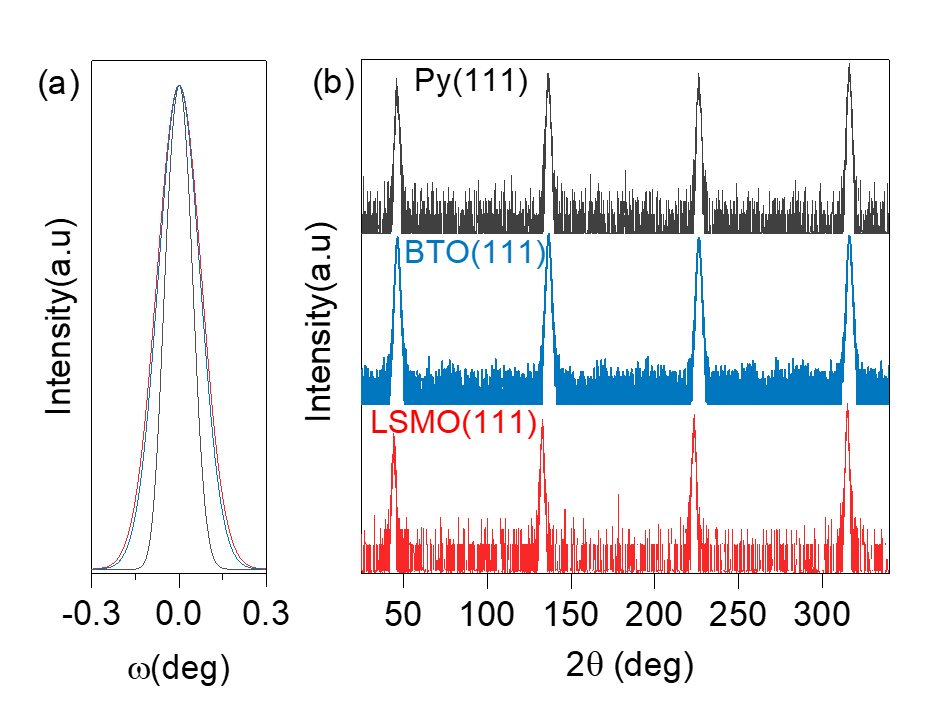}
    \caption{(a) Rocking curve around the (200) Py symmetric plane. The FWHM calculated for the Py/BTO/LSMO(blue), Py/LSMO(red) and Py films(black) is 0.14$^\circ$, 0.15$^\circ$ and 0.09$^\circ$, respectively. (c) XRD $\phi$ scans performed about 111 planes for Py/BTO/LSMO thin film. }
    \label{fig:RCandphi}
\end{figure}
 Fig. \ref{fig:RCandphi} (a) shows XRD rocking curves, performed about the Py (200) crystallographic planes for the heterostructures, yield peaks with a narrow full-width-at-half-maxima FWHM $(0.09^\circ \leq \Delta\omega \leq 0.15^\circ$), confirming the excellent in-plane orientation of the Py layers in all the thin film structures. Fig. \ref{fig:RCandphi} (b) shows the XRD azimuthal $(\phi)$ scans around the (111) asymmetric planes of the Py, BTO, and LSMO layers for the Py/BTO/LSMO thin film. The repeated occurrence of four distinct peaks at intervals of 90$^\circ$ confirms the four-fold cubic symmetry and cube-on-cube epitaxial growth of the layers in the hetero-structure.
\newpage
\section{X-ray photoelectron spectroscopy(XPS)} \label{X-ray photoelectron spectroscopy(XPS)}
\begin{figure}[h]
    \includegraphics[width=1\linewidth]{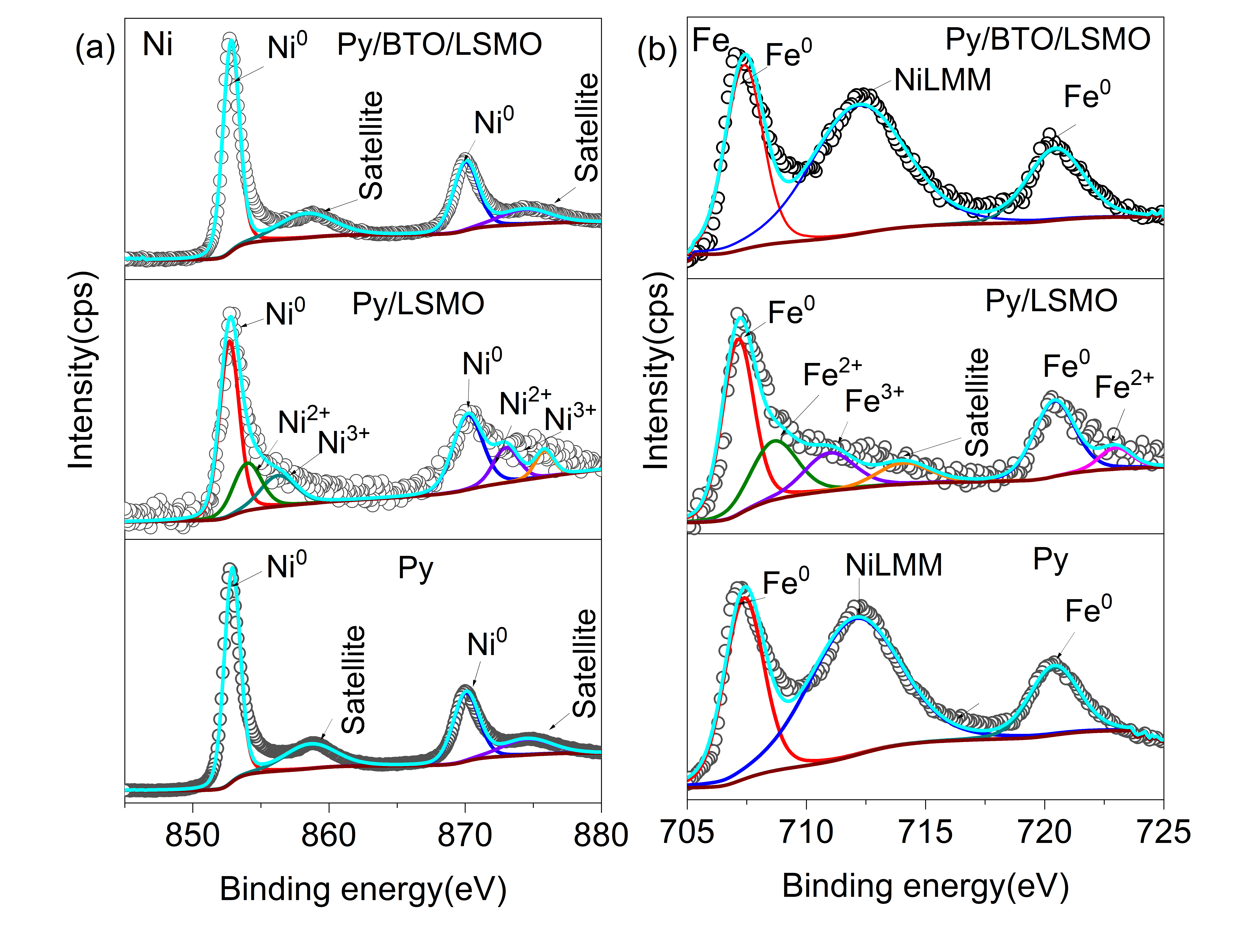}

    \caption{High-resolution XPS spectra of (a) Ni and (b) Fe atoms of the Py layer, for Py, Py/LSMO, and Py/BTO/LSMO heterostructures.}
    \label{fig:xpssupp}
\end{figure}
Fig. \ref{fig:xpssupp} (a) and (b) show the high-resolution XPS spectra of Ni and Fe atoms in the Py layer for the Py, Py/LSMO, and Py/BTO/LSMO heterostructures. The calculated compositions closely align with the nominal composition of Py (Ni$_{0.80\pm0.04}$, Fe$_{0.20\pm0.03}$), confirming the high-quality stoichiometric growth of the heterostructures.
The Ni 2$p$ XPS spectra for all heterostructures display two primary peaks near 853 eV and 870 eV, indicative of the metallic state of Ni \cite{YiFengXPS_1}. In the Py/LSMO heterostructure, additional peaks are observed at 854 eV and 872.9 eV, corresponding to Ni$^{2+}$ \cite{Bodiul_XPS_3, MSALOU_xps_2, Minag_xps_4}, as well as at 856.1 eV and 875 eV \cite{Bodiul_XPS_3, Minag_xps_4}, which are associated with Ni$^{3+}$ alongside metallic Ni. Similarly, the Fe 2p XPS spectra for all heterostructures exhibit peaks at approximately 707 eV and 720 eV, confirming the presence of metallic Fe \cite{YiFengXPS_1}. In the Py and Py/BTO/LSMO heterostructures, the peak around 712 eV originates from Auger NiLMM \cite{Medvedeva_xps_5, YiFengXPS_1}. For the Py/LSMO heterostructure, the Fe 2p spectrum also reveals additional peaks at 709 eV and 723 eV, attributed to Fe$^{2+}$, and at 710.8 eV, attributed to Fe$^{3+}$\cite{MSALOU_xps_2, YAMASHITA_xps_6}, coexisting with metallic Fe. Since all films are coated with a thin Au layer during deposition, the observed additional peaks in XPS spectra of the Py layer in the Py/LSMO heterostructure cannot be attributed to surface oxidation. The simultaneous presence of oxidized (Fe$^{2+}$, Fe$^{3+}$, Ni$^{2+}$, and Ni$^{3+}$) and metallic states in the Fe and Ni XPS spectra highlights charge transfer phenomena between the Py and LSMO layers.

\newpage
\section{In-situ Magneto-optic Kerr effect} \label{In-situ Magneto-optic Kerr effect}
\begin{figure}[h]
    \centering
    \includegraphics[width=0.6\linewidth]{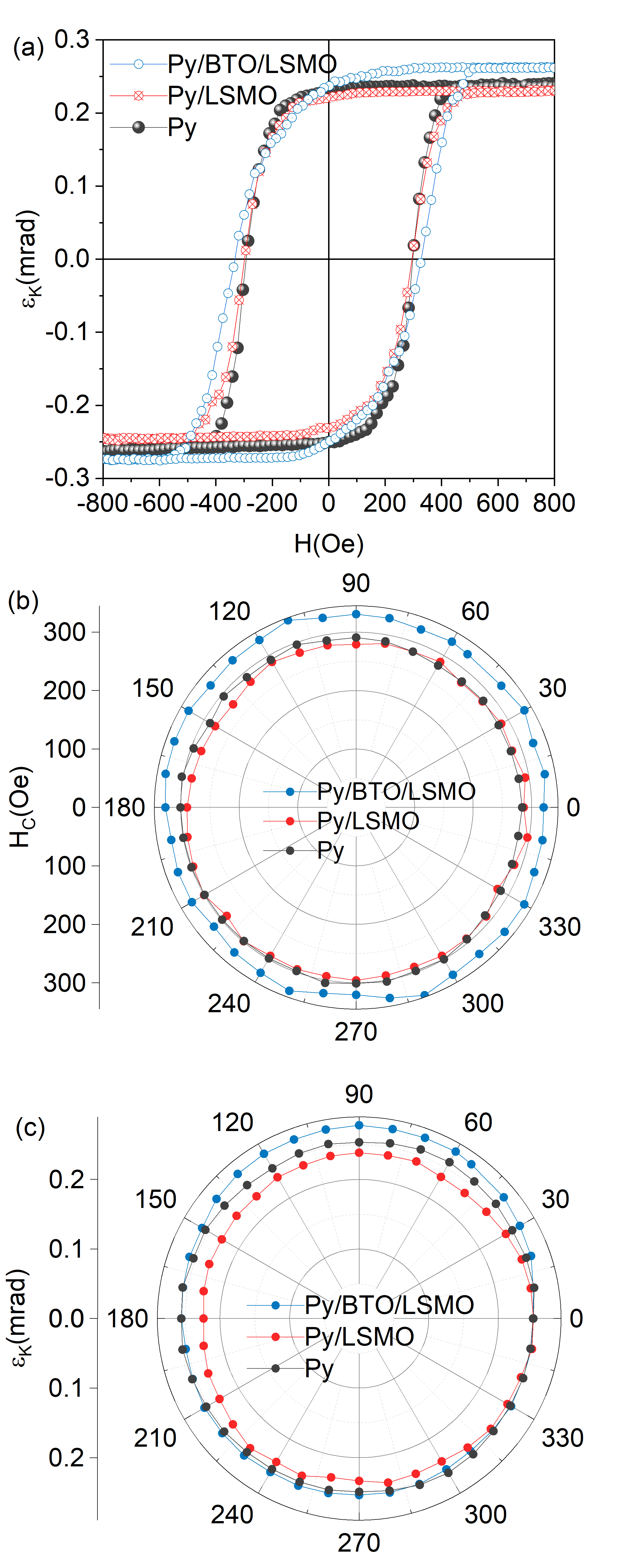}
    \caption{(a) In-plane room temperature longitudinal MOKE $\epsilon_K(H)$ hysteresis loop of Py, Py/LSMO, and Py/BTO/LSMO, respectively. Polar plots of (b) the coercive field (Hc) and (c) saturation Kerr ellipticity, extracted from all the MOKE hysteresis loops of azimuthal scans for $\phi$ = $0^0$ to $360^0$ measured at an interval of $10^0$ for Py, Py/LSMO and Py/BTO/LSMO thin films, respectively. }
    \label{fig:MOKEcoerroom}
  \end{figure}
  Fig. \ref{fig:MOKEcoerroom} displays surface properties of Py, Py/LSMO, and Py/BTO/LSMO heterostructure. The magnetic field in the system applied parallel to the film plane, provides an in-plane magnetic field up to $\pm$ 0.17 T. Longitudinal MOKE hysteresis loops were recorded by applying the magnetic field at various azimuthal angles from $\theta = 0^\circ$ to $\theta = 360^\circ$, with the sample rotated in 10$^\circ$ increments. Fig. \ref{fig:MOKEcoerroom} (a) shows MOKE hysteresis loops for  $\phi$ = 60° for Py, Py/LSMO, and Py/BTO/LSMO thin films. All the thin films show similar MOKE hysteresis which implies that the top Py layers in all the samples are similar in their magnetic properties since the MOKE signal is limited to a thickness of the penetration depth of the laser $(\sim$ 20 nm). Further, the Py top layers do not exhibit any noticeable surface magnetic anisotropy as evidenced by the nearly circular polar plots of the coercive field $(H_C)$ and Kerr ellipticity $(\epsilon_K)$ for the Py, Py/LSMO, and Py/BTO/LSMO thin films as shown in Figs. \ref{fig:MOKEcoerroom} (b) and (c), respectively

\section{Determination of the uniaxial magnetic anisotropy constant} \label{Anisotropy calculation}

The uniaxial magnetic anisotropy constant (K$_u$) for the heterostructures has been calculated by using the equation 
 \begin{equation}
   K_u= \int_{0}^{M_s}(H_{eff}^{out}-H_{eff}^{in})dM    
    \label{eqan}
\end{equation}
where $H_{\rm eff}$ is the externally applied field, $M_S$ is the saturation magnetization, and the subscripts in and out imply in-plane and out-of-plane directions of the applied magnetic field to the film planes. The calculated values of $K_u$ are $4.78\times10^5$ $erg/{\rm cm}^2$ for Py, $1.97\times10^5$ $erg/{\rm cm}^2$ for Py/LSMO and $2.60\times10^5$ $erg/{\rm cm}^2$ for Py/BTO/LSMO.

\section{Determination of OHE, AHE, and THE resistivities from the measured total Hall resistivity } \label{Topological Hall calculation}
The total Hall resistivity can be expressed as
\begin{equation}
   \rho_{xy}(H)= \rho_{xy}^{0}+\rho_{xy}^{A}+\rho_{xy}^{T}= R_0 H+ R_SM+\rho_{xy}^{T}
    \label{eqTHE}
\end{equation}
where the terms $\rho_{xy}^{0}$, $\rho_{xy}^{A}$ and $\rho_{xy}^{T}$  represent the ordinary Hall effect (OHE), anomalous Hall effect (AHE), and topological Hall effect (THE) contributions to the Hall resistivity respectively \cite{Matsuno2016}.
Fig. \ref{fig:THEdata} (a) shows the room temperature total Hall resistivity $(\rho_{xy})$ vs magnetic field curves for Py, Py/LSMO and Py/BTO/LSMO heterostructures. The carrier concentrations calculated from the OHE for the Py film is $1.88(\pm1)\times10^{23}$ cm$^{-3}$, for the Py/LSMO film is $1.02(\pm3)\times10^{23} $ cm$^{-3}$ and for the Py/BTO/LSMO film is $6.77(\pm5)\times10^{22} $ cm$^{-3}$, respectively. Fig. \ref{fig:THEdata} (b) shows the sum of the AHE and THE i.e. $\rho_{xy}^{A+T}=\rho_{xy} (H)-\rho_{xy}^0 $ for the thin films after subtracting the OHE component $\rho_{xy}^0=R_0 H$ calculated by linear-fitting the Hall resistivity at high magnetic fields. At high magnetic field regions in Fig. \ref{fig:THEdata} (b) where Hall resistivity ($\rho_{xy}^{A+T}$) and magnetization saturates $M_{(sat)}$, all the spins align ferromagnetically. This alignment eliminates any THE contribution, leaving behind only the Hall resistivity with the anomalous Hall effect (AHE) from there one can extract  $R_s$ =$\frac{\rho_{xy(sat)}^{A+T}}{M_{(sat)}}$
. By fitting the Hall resistivity using Magnetization, one can extract the AHE response accurately \cite{JihaoYuAM}. After subtracting the AHE ($\rho_{xy}^{A}$) from the sum of AHE and THE ($\rho_{xy}^{A+T}$) we get the THE shown in Fig. \ref{fig:THEdata} (c) ($\rho_{xy}^{T}=\rho_{xy}^{A+T}-\rho_{xy}^{A}$).
  
\newpage
\section{Temperature dependent topological Hall resistivity} \label{Temperature dependent topological Hall resistivity}
\begin{figure}[h]
    \centering
    \includegraphics[width=1\linewidth]{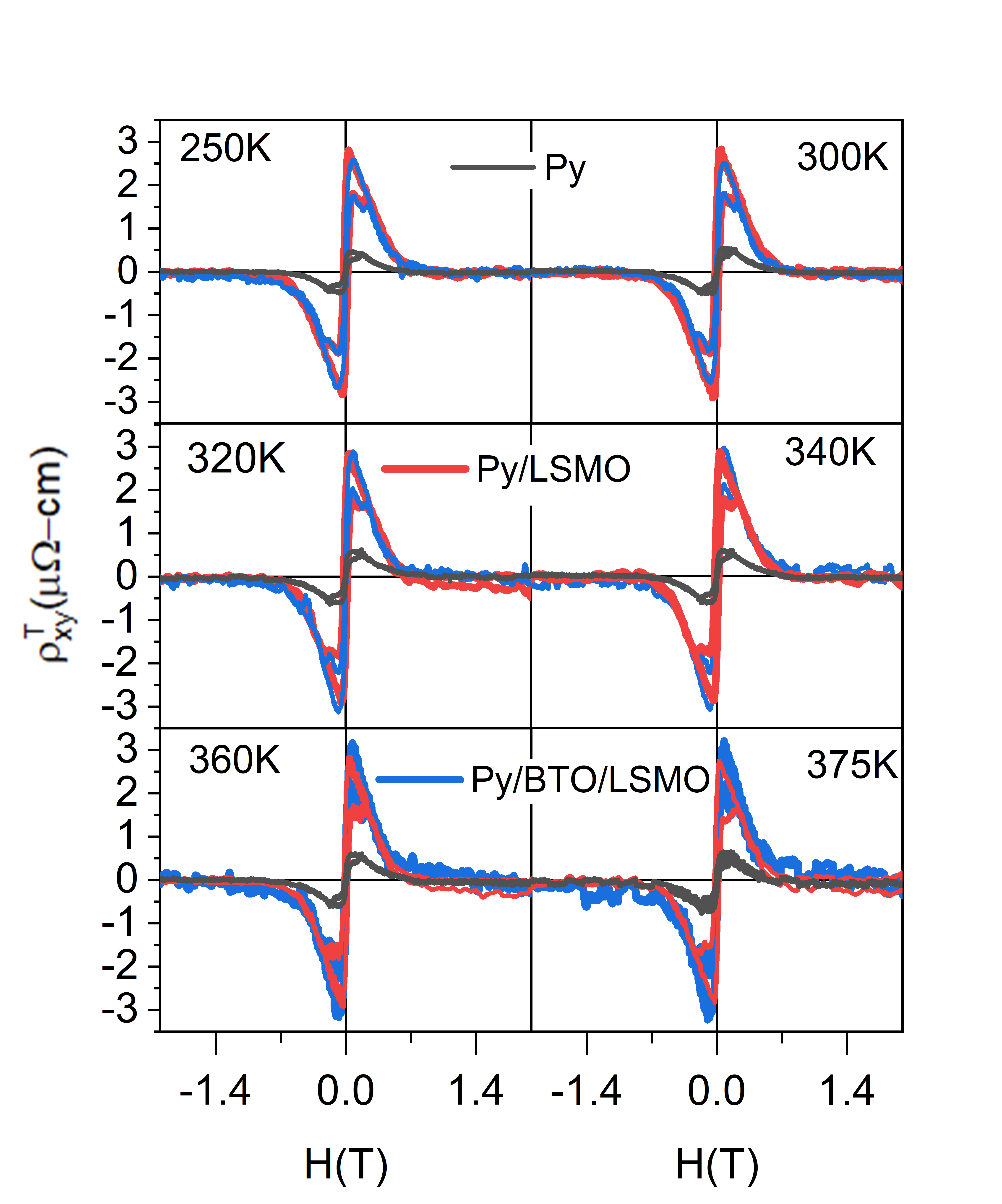}
    \caption{(a)  Temperature-dependent hall resistivity of Py, Py/LSMO, Py/BTO/LSMO heterostructures, showing Py/LSMO has maximum Hall resistivity till 320 K, between 320 K to transition temperature of LSMO (340 K) both has nearby same resistivity, above the transition temperature of LSMO, Py/LSMO Hall resistivity decreases but in Py/BTO/LSMO it increases, due to BTO intrinsic polarization }
    \label{fig:tempHall}
\end{figure}
Fig. \ref{fig:tempHall}  shows the field dependence THE ($\rho_{xy}^T$)  for Py (Black), Py/LSMO (Red), Py/BTO/LSMO (Blue) heterostructures at different temperatures from 250 K to 375 K. At all the temperatures, the enhanced THE is observed in the Py/LSMO and Py/BTO/LSMO heterostructures indicating the stability of the observed effect. 

\newpage
\section{Magnetic Force Microscope (MFM)} \label{Magnetic Force Microscope_suppl (MFM)}
\begin{figure*}[t]
    \includegraphics[width=1\linewidth]{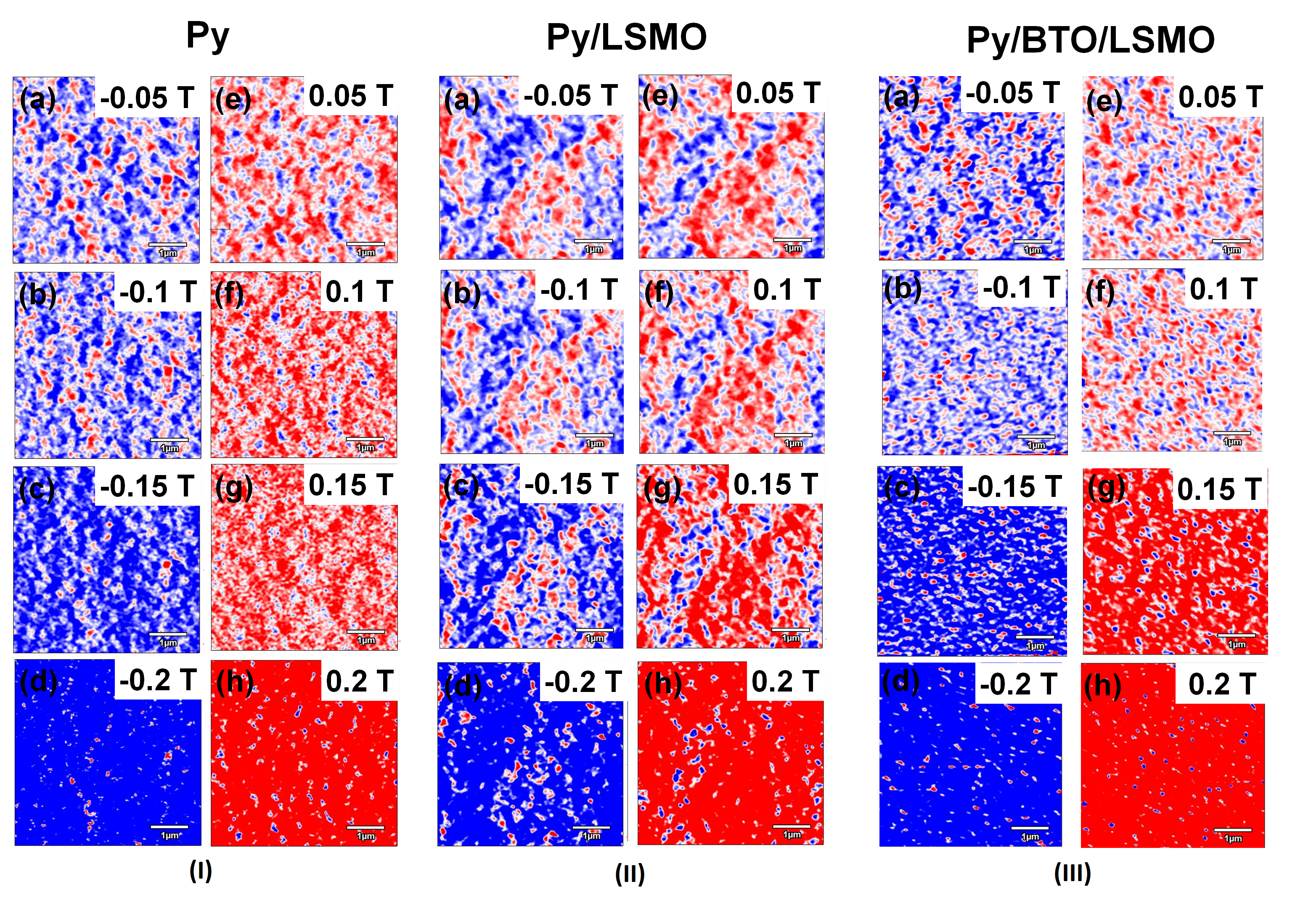}
    \caption{    Effect of external magnetic field-dependent MFM images varying field in the range of 0.05 T to 0.2 T at a step of 0.05 T field for (I) Py, (II) Py/LSMO, and (III) Py/BTO/LSMO  films at (a) -0.05 T, (b) -0.1 T, (c) -0.15 T, (d) -0.2 T, (e) 0.05 T, (f) 0.1 T, (g) 0.15 T, (h) 0.2 T. The scale bar at the bottom of the images represents 1$1\mu m$ }

    \label{fig:MFMsupplemetry}
\end{figure*}
Fig.~ \ref{fig:MFMsupplemetry} (I, II, and III) presents magnetic field-dependent MFM images for Py, Py/LSMO, and Py/BTO/LSMO films, respectively. These images were obtained by varying the magnetic field in positive ($+z$) and negative ($-z$) directions within a range of $-$0.2 T to +0.2 T, with increments of +0.05 T. The contrast between the red and blue regions corresponds to magnetization along the $+z$ and $-z$ directions, respectively. At the same time, white areas indicate spins oriented in the plane of the thin film, resulting in diminished magnetic signals.
The MFM images (a, b, c, and d) reveal that as the magnetic field strength increases, the magnetic domains expand, aligning progressively along the field direction. This trend continues; however, until $-$0.2 T, all domains should have turned blue, but we have observed red regions revealing the emergence of small, skyrmion-like structures for the Py (I), Py/LSMO (II), and Py/BTO/LSMO (III) heterostructures. 
Comparing panels (a, b, c, and d) with (e, f, g, and h) for all heterostructures, it was observed that reversing the field direction from $-$0.05 T to $-$0.2 T produces features analogous to those seen in the range of 0.05 T to 0.2 T discussed in the main text.
\newpage

\bibliography{apssamp}

\end{document}